# An Integrated Mobile Application for Enhancing Management of Nutrition Information in Arusha Tanzania.

Neema Mduma
School of Computation and Communication Science & Engineering
Nelson Mandela African Institution of Science and Technology
Arusha, Tanzania

Khamisi Kalegele
School of Computation and Communication Science & Engineering
Nelson Mandela African Institution of Science and Technology
Arusha, Tanzania

*Abstract* — Based on the fact that management of nutrition information is still a problem in many developing countries including Tanzania and nutrition information is only verbally provided without emphasis, this study proposes mobile application for enhancing management of nutrition information. The paper discusses the implementation of an integrated mobile application for enhancing management of nutrition information based on literature review and interviews, which were conducted in Arusha region for the collection of key information and details required for designing the mobile application. In this application, PHP technique has been used to build the application logic and MySQL technology for developing the back-end database. Using XML and Java, we have built an application interface that provides easy interactive view.

## I. INTRODUCTION

The mobile technology has been the most fastest growing media technology used in the healthy sector in Tanzania in recent years compared to other media technologies [1]. This technology directly targets the general public through engaging users in health related activities, and thereby improving accessibility to quality health information, health services, and encouraging user behavior that involves seeking preventive health solutions [2]. The wide spread of mobile phones has led to significance increase in mobile applications for providing access to various information that are needed by the community. Mobile applications have been designed to run on mobile devices and allow users to interact with service providers. Our proposed system is in the form of an integrated mobile application, which is designed to enhance management of nutrition information.

The proposed mobile application will be integrated with the existing health centre system. The health centre system used is Open MRS. The proposed system will allow nutrition practitioners to send information and recommendations to the targeted user. In this aspect, the user will be able to access nutrition information and request any other nutrition related details or seek advice when necessary. The application will also provide reminders so as to notify the user on necessary events such as clinic visit for vitamin A supplements. In this application, nutrition tips will be generally provided and available for all users, and the users will be able to view nutrition tips and request for new tips based on their concern and nutrition practitioners will respond to the request accordingly. In responding to the nutrition tips enquiries, nutrition practitioners' profile will be specified so as to show the validity of the tip. In this application, the researcher will be able to generate nutrition reports based on provided information and the administrator will monitor the overall activities of the system and be responsible for user approval. The application will allow user interaction whereby the authorized user will be able to view the historical recommendation and request assistance when needed. The system will be user interactive and support two way flow of nutrition information.

## II. METHODOLOGY

The requirement gathering was conducted in Arusha region. The method used in this study based on qualitative research methods such as literature review and interviews whereby casual talks were conducted for the collection of information. Through the interviews, we interacted with the nutrition practitioners together with prenatal and post-natal mothers and collected data relevant for specifying the requirements for developing the mobile application.

## III. REQUIREMENT SPECIFICATION

This study involves both functional and non-functional requirements. The functional requirements for developing this mobile application covers the issues of recommendation as set of nutritional information that are suggested by nutrition practitioners to the user based on the user's described information and nutrition tips as the set of nutritional information concerning nutrition improvements added by nutrition practitioners for the user. The functional requirements also include a reminder as notification provided to users based on necessary nutrition events and reports that are generated by researcher based of nutrition information provided. The non-functional requirements of the system cover the issues of maintainability, operability, performance and security of the system.

## IV. DESIGNING THE PROPOSED SYSTEM

In this study, the design part was illustrated on two major parts using Data Flow Diagram (DFD). Fig.1 shows administration management data flow diagram and Fig.2 shows tips and recommendation management data flow diagram.

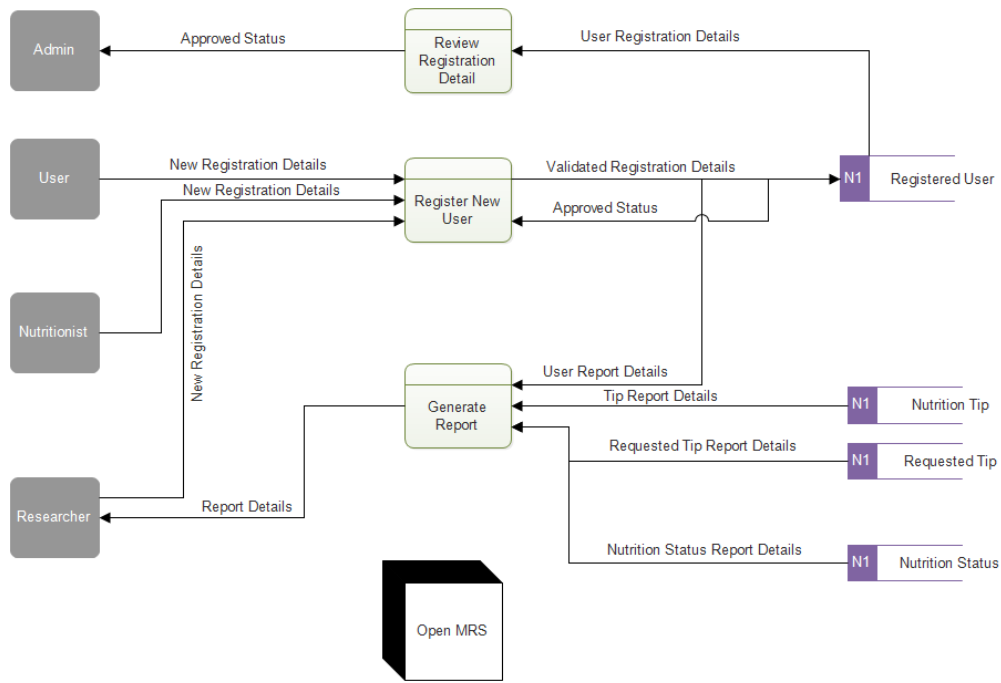

Fig. 1 Administration management data flow diagram

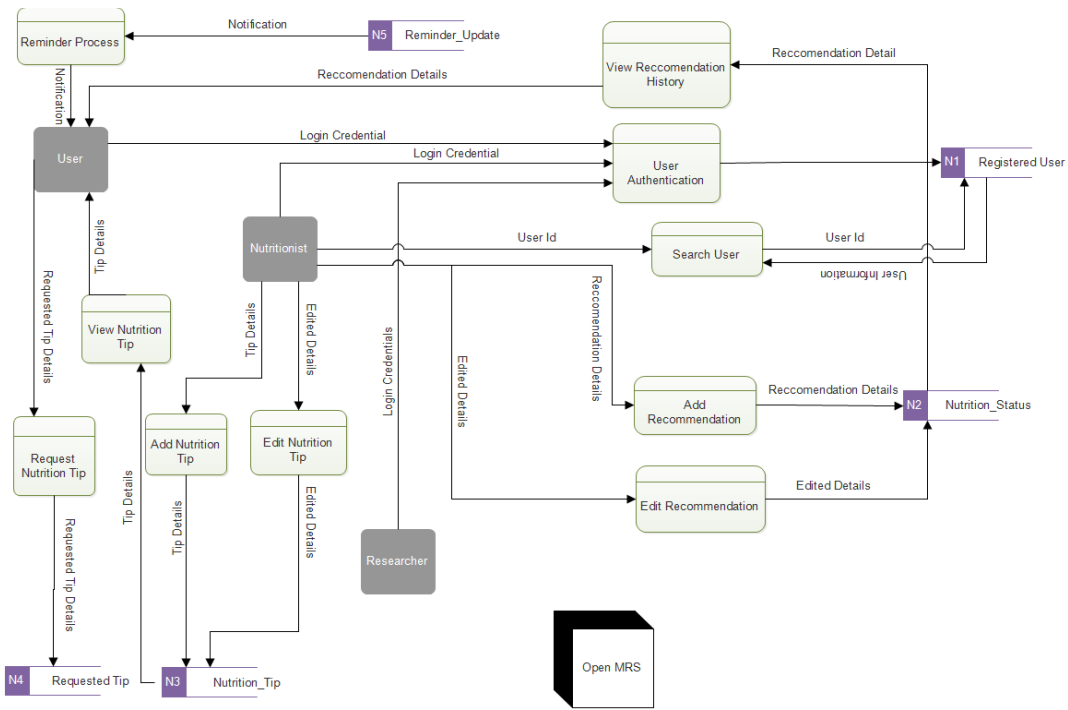

Fig. 2 Tips and Recommendation management data flow diagram

To accomplish the implementation part for the proposed application, a model adopted from SDLC has been chosen for developing a successful information system. The software development life cycle (SDLC) is a framework that defines the tasks performed at each step in the software development process. It consists of a meticulous plan that describes the processes for developing, maintaining, replacing and altering the specific software. The SDLC defines the method for software quality enhancement and the overall development process [3]. To make the complete product to deliver faster, we decided to use the Rapid Application Development (RAD) model.

*A. Rapid Application Development*

The RAD is a model designed to facilitate much faster software development and provides higher quality results compared to the traditional lifecycle; this model delivers faster and higher quality product [4]. In this study, we preferred to use RAD as it proved to be successful tool for developing our mobile application. Fig. 3 shows the Rapid Application Development model of our system.

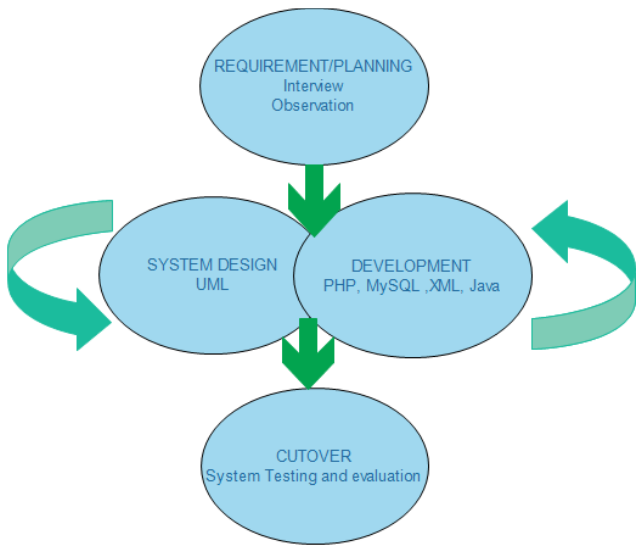

Fig. 3 The System's Rapid Application Development Model

*B. Mobile Application*

Mobile application is a type of application software that takes advantage of the mobile technology, and it can be used with any other technology apart from mobile phones [5]. The numerous functions and services offered prompt the extensive use of the mobile applications. In this paper, we use android mobile application in order to distinguish with other Unstructured Supplementary Service Data (USSD) applications that provide limited information and don't support storing of provided information. The reason is to provide two way flow of information by supporting interaction and allow access of large amount of information.

*1) Why choose Android:* Android is one of the most powerful and flexible open source platforms and its increasingly becoming popular. There are no licensing fees; this increases preference of many developers. In this study, we preferred developing mobile application supported by this operating system with consideration of market terms. The growth of mobile devices such as mobile phones is a worldwide phenomenon with mobile phone ownership outstripping computer ownership in many countries. Also there is an increase in smart phones, which created a growth market for advanced mobile applications [6].

*C. PHP*

In developing the mobile application, we used the Hypertext Pre-processor (PHP) because this is one of the server-sided languages widely-used in software development and is an open source scripting language that we found appropriate for developing our system.

*1) Why use PHP:* PHP was preferred in this development study because, first it is simple and thus easy to learn. It efficiently runs on the server side and its codes runs faster due to the fact that it runs in its own memory space so it has a fast loading time. The PHP has tools that are open source software, and thus are freely available for use. Furthermore, it is flexible for database connectivity and it supports a wide range of databases. Additionally, the PHP can connect to a number of databases, but MySQL is the most commonly used as it can also be used at no cost [7].

In addition, PHP is compatible with almost all servers and its security features allow many functions to protect users against certain attacks. This language runs on various platforms such as Android, Windows and so many others.

*D. MySQL*

MySQL is one of the database systems that run on a server and uses the standard Structured Query Language (SQL). It is easy to use, reliable and it runs very fast. In this study, we used MySQL Database so as to enable the cost-effective delivery of reliable and high-performance application. The data in a MySQL database are stored in tables and offers a flexible programming environment [8]. Database systems are vital in computing and can be used as standalone utilities or as part of other applications.

*1) Why use MySQL:* The MySQL database server provides the ability to handle applications that are deeply embedded and offers platform flexibility; this is a MySQL stalwart feature. It allows customization so it is easy for a programmer to improve the database server by adding unique features.

MySQL has been used by many database professionals due to the unique storage-engine architecture that allows configuration of the database server remarkable end results performance in particular applications.

Apart from that, MySQL offers a variety of unique high-availability database server options ranging from high-speed master/slave replication configurations, specialized cluster servers offering instant failover, to third party vendors. So it provides high availability for programmers to rely on it.

MySQL protects data through its outstanding security features; it has powerful mechanisms, which ensures that access to the database server is possible only to authorized users and other users are limited to the client machine level.

MySQL also has granular object privilege framework, which ensures that users can only see what they are supposed to see. Another important feature is that it has powerful data encryption and decryption functions, which protects sensitive data from unauthorized users. Secure Shell (SSH) and Secure Sockets Layer (SSL) are provided to ensure safe and secure connections. It also provides backup and recovery utilities so as to allow complete logical and physical backup, and also full and point-in-time recovery.

MySQL offers full support needed for development of applications and developers can get all they required for developing information systems that are based on databases. [9].

*E. XML*

Extensible Markup Language (XML) is designed to describe data [10]. This language is used as a medium for carrying information independently from the involved software and hardware of the information system. Through the XML, you can create information formats and structured data can be shared electronically. XML data is self-describing, which means the data and its structure are embedded replacing the need for pre-building the structure for storing the data when it arrives. XML allows sharing of information in a consistent way due to its simpler format [11].

*1) Why XML:* XML has good features for storing and transmitting information, which simplifies data storage and sharing. This language is useful in accurately describing and identifying information without mistake so as to allow information to be understood [12]. Standardized description and control of particular types of document structure is possible in XML. It provides messaging systems' common syntax to facilitate information exchange between applications. In this study, we decided to use XML because it is free so we don't need to pay and it is easier to upgrade without losing data.

*F. Java*

Java is a programming language and computing platform that is designed to support many applications to work. This language is fast, secure, and reliable so as to ensure developers about performance, stability and security of the developed application [13].

*1) Why Java:* In this study, we decided to use Java because it is platform independent so applications can run on many different types of devices such as computers and even mobile phones. Java is essentially made up of objects, which are programming elements, and therefore it is object-oriented [14]. This language is very simple, so it is easier for the developer to engage it in application development.

V. RESULTS

An integrated mobile application has been developed as a result for enhancing management of nutrition information and integration with existing health system as declared in this study. Results show the system interface that was developed by using XML and Java so as to allow user interaction with the system.

*A. System Interface*

Designing an interface is described as the process of developing a method in a system to connect and communicate so as to allow exchange of information. This acts as a channel of communication between user and application. Interface design focuses on anticipating what users might need to do and ensuring that the interface has elements that are easy to access, understand, and use to facilitate those actions [15].

*1) Interface for mobile application:* This section provides some of the developed interface for this application. First, the system administrator will register the users by approving their registration requests as no one can use the system without registration. Users will be using mobile phones to access this application. The application interface is presented in the Fig. 4 below.

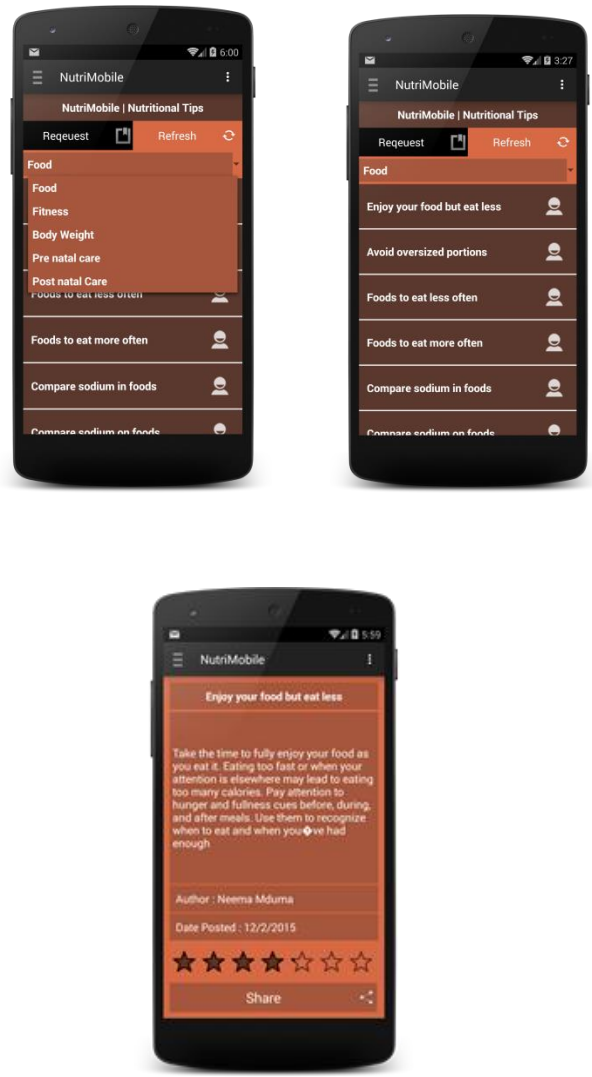

Fig. 4 System interfaces

## VI. CONCLUSION

This study was on developing an integrated mobile application for managing nutrition information in Tanzania. The system development used various methods and materials, which were determined after the design process discussed in this paper, and which culminated into development of a mobile application for management of nutrition information. Mobile phones were chosen as the tool to manage nutrition information so as to allow interaction without time and place limitations due to ownership issues. After registration, only authorized users will be able to access the information. The system administrator is the one responsible for the approval of user registration and this will provide security. All nutrition information is provided by nutrition practitioners and the system will allow sharing of that information via social networks. The user will be reminded in case of any necessary event concerning nutrition and clinic visits so as to increase efficiency. On the other hand, the user can request for nutrition information and nutrition practitioners will respond accordingly.

REFERENCES

[1] H. Cole-Lewis and T. Kershaw, "Text messaging as a tool for behavior change in disease prevention and management," *Epidemiologic Reviews*. 2010.

[2] Rick van Genuchten, Wouter Haring, Daan van Kassel, and Kaoutar Yakubi, "Mobile phone use in Tanzania," *TAN2012 Mark. Res.*, 2012.

[3] K. Schwaber, "Software Development Life Cycle (SDLC)," *Tutorials point simple easy Learn.*, 2001.

[4] M. A. Hirschberg, "STN 2-1 Topic: Rapid Application Development (RAD): A Brief Overview," 2015. [Online]. Available: https://www.mendeley.com/library/viewer/?fileId=6f1c4f30-d7da-50d9-67a7-2c02747d1f83.

[5] J. Muthee and N. Mhando, "AMDI-BBC-summary-report," *African Media Dev. Initiat.*, 2006.

[6] R. Meier, *Professional Android 2 Application Development*. 2010.

[7] Pitts Monica, "5 Reasons Why PHP Is a Great Programming Language | Mayecreate Design," 2015. [Online]. Available: http://www.mayecreate.com/2013/12/5-reasons-php-great-programming-language/. [Accessed: 16-May-2015].

[8] Heng Christopher, "What is MySQL? What is a Database? What is SQL? (thesitewizard.com)," 2010. [Online]. Available: http://www.thesitewizard.com/faqs/what-is-mysql-database.shtml. [Accessed: 16-May-2015].

[9] MySQL, "MySQL :: Top Reasons to Use MySQL," 2015. [Online]. Available: https://www.mysql.com/why-mysql/topreasons.html. [Accessed: 16-May-2015].

[10] Hall Marty and Brown Larry, "Microsoft PowerPoint - XML.ppt - xml.pdf," 2001. [Online]. Available: http://www.cs.toronto.edu/~mlou/csc309/xml.pdf. [Accessed: 17-May-2015].

[11] Rouse Margaret, "What is XML (Extensible Markup Language)? - Definition from WhatIs.com," 2001. [Online]. Available: http://searchsoa.techtarget.com/definition/XML. [Accessed: 17-May-2015].

[12] Tutorialspoint, "xml_tutorial.pdf," 2014. [Online]. Available: http://www.tutorialspoint.com/xml/xml_tutorial.pdf. [Accessed: 17-May-2015].

[13] Eck David, Hobart, and Colleges William, "() - javanotes6-linked.pdf," 2014. [Online]. Available: http://math.hws.edu/eck/cs124/downloads/javanotes6-linked.pdf. [Accessed: 17-May-2015].

[14] Lowe Doug and Burd Barry, "What Is Java, and Why Is It So Great? - For Dummies," 2007. [Online]. Available: http://www.dummies.com/how-to/content/what-is-java-and-why-is-it-so-great.html. [Accessed: 17-May-2015].

[15] Usability, "User Interface Design Basics," May 2014.